\documentclass{bjour}

\newcommand{\be}{\begin{equation}}
\newcommand{\ee}{\end{equation}}
\newcommand{\bea}{\begin{eqnarray}}
\newcommand{\eea}{\end{eqnarray}}
\def\fct#1{\mathop{\rm #1}}	
\def\im{\fct{Im}\:}
\def\re{\fct{Re}\:}
\newcommand{\Eq}{Eq. \ref}
\newcommand{\Fig}{Fig. \ref}
\newcommand{\Cz}{\mbox{\boldmath $C$}}

\def\eps{\varepsilon}
\def\<{\langle} 				
\def\>{\rangle} 				

\begin{document}


\authorrunninghead{Mandelshtam and Neumaier}
\titlerunninghead{pseudo-time Schr\"odinger equation}
\title{Further generalization and numerical implementation of pseudo-time Schr\"odinger equations for quantum scattering calculations
} 
\received{April 11, The Journal of Theoretical and
Computational Chemistry (http://www.worldscinet.com/jtcc/jtcc.shtml)}

\author{Vladimir A. Mandelshtam$^1$ and Arnold Neumaier$^2$}

\affil{$^1$ Chemistry Department, 
University of California at Irvine, Irvine, CA 92697, USA; 
email: mandelsh@uci.edu}

\affil{$^2$ Institut f\"ur Mathematik, Universit\"at Wien
Strudlhofgasse 4, A-1090 Wien, Austria;\\ email: Arnold.Neumaier@univie.ac.at; 
WWW: http://www.mat.univie.ac.at/$\sim$neum/}

\date{\today}

\abstract{  
We review and further develop the recently introduced numerical approach
[Phys. Rev. Lett. {\bf 86}, 5031, (2001)]
for scattering calculations based on a so called {\it pseudo-time Schr\"odinger
equation}, which is in turn a modification of the damped Chebyshev
polynomial expansion scheme [J. Chem. Phys. {\bf 103}, 2903, (1995)].
The method utilizes a special energy-dependent form for the absorbing
potential in the time-independent Schr\"odinger equation, in which the
complex energy spectrum is mapped inside the unit disk $E_k\to u_k$, where
$u_k$ are the eigenvalues of some explicitly known sparse matrix
$U$. Most importantly for the numerical implementation, all the
physical eigenvalues $u_k$ are the extreme eigenvalues of $U$ (i.e., $u_k\approx 1$ for resonances and
$u_k=1$ for the bound states), which allows one to extract these
eigenvalues very efficiently by {\it harmonic inversion} of a
pseudo-time autocorrelation function $y(t)=\phi^{\rm T} U^t \phi$
using the {\it filter diagonalization method}. The
computation of $y(t)$ up to time $t=2T$ requires only $T$ sparse real
matrix-vector multiplications. We describe and compare different
schemes, effectively corresponding to different choices of the energy-dependent
absorbing potential, and test them numerically by calculating
resonances of the HCO molecule. Our numerical tests suggest an optimal
scheme that provide accurate estimates for most resonance states
using a single autocorrelation function.
} 



\begin{article}


\noindent 
{\bf Introduction.}
In this paper we present a detailed description and further
generalization of the methodology developed in the preceding works
\cite{cheb,filter2_JCP,cross,LiG,res_prl} for the efficient numerical solution of
the quantum scattering problem associated with the time-independent
Schr\"odinger equation
\be\label{ab.1}
   (H\psi)(r)=E\psi(r).
\ee

\Eq{ab.1} may possess {\it bound states}
with real energies $E$ and wavefunctions
$\psi(r)$ exponentially localized in space, and {\it resonance
states} 
(Siegert states \cite{siegert}) having complex energies 
with $\im E\leq 0$. The latter behave like 
bound states in some compact subset $\Omega$ of the configuration
space, but eventually grow exponentially
outside of $\Omega$, due to the outgoing asymptotic boundary conditions.

The corresponding bound state problem is conceptually simple and 
very well understood. 
The numerical solution of (\ref{ab.1}) at dissociation energies is
much more difficult as it requires the solution of a boundary value
problem. 
One can avoid the latter by the use of a
so-called {\it optical (or absorbing)
potential} $W(r)$ with $\im W(r)\leq 0$ 
(in the sense that $i(W-W^*)$ is positive semidefinite,
where $^*$ denotes conjugate
transposition), that
vanishes for $r\in\Omega$ and smoothly grows outside
$\Omega$. Numerically, this has negligible effect on 
the scattering solutions $\psi(r)$ of 
\be\label{ab.2}
   (H\psi)(r)=(E-W(r))\psi(r)
\ee
inside $\Omega$, i.e.,  the physically relevant region, and damps them
outside $\Omega$ \cite{Jolicard}. In other words, the complex
absorbing potential 
forces the resonance solutions to behave like bound states everywhere without
significantly affecting the energies $E$. In this framework
the physically relevant part of the system is, therefore, 
dissipative and satisfies (\ref{ab.1}) only for $r\in\Omega$.
Moreover, a general multichannel scattering problem 
can be considered with a numerically 
convenient form of $W(r)$, independent of
the choice of coordinate system. The price for these benefits is that
the originally hermitian problem becomes non-hermitian; 
but it is generally still complex symmetric.

Although to satisfy  $\im E\leq 0$ one only needs
$\im W\leq 0$, traditionally one simply uses a negative imaginary
potential $W=-i\Gamma$ and gets 
the nonhermitian eigenvalue problem $(H-i\Gamma)\psi=E\psi$. The latter is
generally much easier to handle numerically 
than the boundary value problem (\ref{ab.1}). 
As we already saw in \cite{cheb,res_prl}, energy-dependent choices
$W=W_E(r)$ are particularly useful.  

The introduction of the absorbing potential leads to the {\it damped
Green's function} \cite{NeuhBaer,SeidMill,RisM}
\be\label{ab.GW}
G_W(E):=(H-E+W_E)^{-1}.
\ee
We assume that under suitable conditions on $W_E(r)$, 
similarly to the traditional case $W=-i\Gamma$ \cite{RisM},
$G_W(E)$ converges for any real
$E$ (and also for $\im E\ge 0$) 
weakly to the ordinary Green's function
\[
G(E)=\lim_{\eps\downarrow 0} (H-E-i\eps)^{-1}.
\]
Practically, one usually needs to evaluate only certain matrix elements 
$\phi^{\rm T}G(E)\psi$, the basic numerical 
objects of quantum physics, 
from which most other quantities of interest (scattering amplitudes, 
reaction rates, etc.) can be computed (see, e.g.,
refs.~\cite{NeuhBaer,SeidMill,TanWeeks,Kourivar}). 
If both $\phi$ and $\psi$ have support in $\Omega$, they are well
approximated by $\phi^{\rm T}G_W(E)\psi$.

Unfortunately, for very large systems with high density of states one 
may encounter numerical difficulties when trying to diagonalize a large
nonhermitian matrix $H'=H+W$ or solve the linear system 
$(E-H')\phi(E)=\psi$ at many values of $E$ using
general iterative techniques for nonhermitian matrices.
For instance, the Krylov subspace algorithms, such as the 
Lanczos diagonalization procedure, usually converge well for
the extreme eigenvalues of $H'$, while numerical problems may  
occur for interior complex eigenvalues of $H'$ in the dense spectral
regions, requiring more sophisticated schemes. 
As such a new technique called PIST was recently
introduced by Poirier and Carrington
\cite{poirier}. The PIST method is based on a very efficient
preconditioning within a QMR-Lanczos framework for iterative
diagonalization of large and sparse DVR Hamiltonians with
complex absorbing potentials.
It is
also appropriate to mention the 
time-dependent approach based on
solution of the time-dependent Schr\"odinger equation, 
\be\label{ab.tdse}
\phi(t)=e^{-itH'}\phi(0),
\ee
which is widely used because of its simplicity, 
and can also be viewed as a Krylov subspace method with Krylov
vectors 
\be\label{ab.tdse1}\phi(t)=U^t\phi(0)\ee 
generated by the powers of the evolution
operator $U=e^{-iH'}$. The
bound state eigenvalues $\lambda_k=e^{-iE_k}$ of $U$ appear at the unit
circle; the resonance eigenvalues near the unit circle,
$|\lambda_k|\sim 1$ , and
satisfy $|\lambda_k|\leq 1$. That is, for a general initial state
$\phi(0)$  all the physically important
states $\psi_k$ significantly contribute to $\phi(t)$, because they
all correspond to the {\it extreme eigenvalues} of $U$ (for which the
relative weights defined by $|\lambda_k|^t$ are significant). 
This makes $\phi(t)$ a
convenient basis for performing the spectral analysis of $H'$.
For example, one can generate a time-correlation function
\be
C(t):=\psi^{\rm T}\phi(t).
\ee
Because the {\it time signal} $C(t)$ satisfies the form 
\be\label{ab.HIP}
C(t):=\sum_{k=1}^K d_k \lambda_k^t
\ee
with the weights $d_k=\psi^{\rm T}\psi_k\psi_k^{\rm T}\phi$,
one can use Fourier spectral analysis to extract the desired
eigenvalues $E_k$. Furthermore, the recently developed
{\it Filter Diagonalization Method} (FDM) \cite{Neuh95,filter2_JCP}
(see also refs.~\cite{RRT_JPC,XFT_JMR} on other related
superresolution methods of spectral analysis) 
to solve the {\it harmonic inversion problem} (\ref{ab.HIP}) with
the unknowns $\{d_k,\lambda_k\}$ generally leads to an
enormous resolution enhancement, thus significantly reducing the required
propagation time, as well as the overall numerical work.
At first glance the time-dependent framework seems nearly optimal
as the time correlation functions can be generated at low cost by
various iterative techniques, e.g., the split-operator method \cite{split}. 
The
difficulty, however, arises when both the density and the
number of states are very high, in which case
the time domain data $C(t)$ must be very accurate at very long times
in order to provide the adequate conditions for, e.g., FDM. 
Unfortunately, this requirement is very hard to
satisfy, as it is difficult to accurately evaluate the matrix
exponential  $e^{-itH'}$ for a non-hermitian operator $H'$ at very large
values of $t$.

Apparently, for a quantum system with Hamiltonian operator that is not
explicitly time-dependent, there is nothing special about the
time-dependent Schr\"odinger equation, except that it provides a
convenient framework for both thinking and devising various numerical
techniques to solve the time-independent problem, for example, those
based on processing the time correlation functions. As such we can consider
alternative dynamical schemes having the convenient structure of
Eqs.~\ref{ab.tdse1}-\ref{ab.HIP},
albeit with a pseudo-evolution operator $U$, somehow related to the
underlying Hamiltonian $H$, but whose action on a vector can
be evaluated easily.
To this end the analogy between the
standard time evolution  and the Chebyshev recursion ,
\be\label{ab.cheb}
\phi(t)=2H\phi(t-1)-\phi(t-2)
\ee
with initial conditions $\phi(0)=\phi,\ \phi(1)=H\phi$,
or, more generally, damped Chebyshev
recursion \cite{cheb}, 
\be\label{ab.dampcheb}
\phi(t)=2DH\phi(t-1)-D^2\phi(t-2)
\ee
with initial conditions $\phi(0)=\phi,\ \phi(1)=DH\phi$ 
and damping operator $D$ ($D(r)=1$ inside $\Omega$ and  $D(r)\le 1$, outside), was noticed and
explored previously
\cite{SpProj,ChenGuo,filter2_JCP,Gray_real}.
In ref. \cite{res_prl} it was
shown explicitly that a variant of \Eq{ab.dampcheb},
\be\label{ab.dampcheb1}
\phi(t)=2DH\phi(t-1)-D\phi(t-2)
\ee
with initial conditions $\phi(0)=\phi,\ \phi(1)=0$, 
can be written in the familiar form
(\ref{ab.tdse1}) with some effective evolution operator $U$. The corresponding {\it pseudo-time Schr\"odinger
equation} allows one to reap all the benefits of the time-dependent 
methods 
without having to deal with the standard time-dependent 
Schr\"odinger equation (\ref{ab.tdse}) 
involving nonhermitian Hamiltonian; it only requires the evaluation
of a single $H$-matrix-vector product per time step and avoids
the use of complex arithmetic, even when the absorbing potential is  
implemented.

As is well known, a (physical time) autocorrelation function at time 
$2t$ can be computed by solving the time-dependent Schr\"odinger
equation up to time $t$, since one can use
\[
C(t):=\phi^{\rm T}U^{2t}\phi=(U^t\phi)^{\rm T}(U^t\phi),
\]
assuming only the complex symmetry of the evolution operator $U$,
which is always easy to achieve, both with absorbing potential or 
without.
For the Chebyshev autocorrelation function 
\be \label{ab.autocor} y(t):=\phi^{\rm T}\phi(t),\ee 
where the vectors $\phi(t)$ are generated by \Eq{ab.cheb},
a factor of two saving is also well known 
(see, e.g., the discussion in ref.~\cite{filter2_JCP}):
\[
y(2t+p)=2\phi(t)^{\rm T}\phi(t+p)-y(p)
\]
with $p=0,1$.
When the damped scheme
(\ref{ab.dampcheb}) is implemented, 
this recipe does not provide the correct autocorrelation function
as defined by \Eq{ab.autocor}. Nevertheless, it was tried by Li
and Guo \cite{LiG}. The resulting doubled sequence, when processed
by FDM, gave approximately correct resonance energies and widths.
In ref.~\cite{res_prl} starting with the
recursion formula (\ref{ab.dampcheb1}) we derived an exact doubling
scheme for the corresponding autocorrelation function (\ref{ab.autocor}):
\[
y(2t+p)=\phi(t)^{\rm T}\phi(t+p)-\phi(t+1)^{\rm T}D^{-1}\phi(t+1+p)
\]
with $p=0,1$.

We note that the damping operator $D$ in \Eq{ab.dampcheb}, 
as well as in \Eq{ab.dampcheb1}, effectively leads to an energy-dependent 
complex absorbing potential $W_E(r)$, which may have a small or large real 
part relative to its imaginary part, depending on the particular
recursion formula and the
energy $E$.
Usually, having too large $\re W_E(r)$ is not desirable. 
Unfortunately, within the two frameworks one has little control over
this circumstance.
In the present paper, we re-derive and extend the above results by  
using a more flexible form for the absorbing potential, which can be
adapted easily. The resulting pseudo-time Schr\"odinger equation is
shown to be very convenient for calculating various dynamical
properties, such as resonance parameters or matrix elements of the
Green's function. The latter is carried
out by generating the pseudo-time cross-correlation functions followed
by their inversion. A numerical example demonstrates the validity of
the theory and compares different damping
schemes, suggesting an optimal scheme, that leads to accurate results
with minimal computational effort.

\section*{Quadratic eigenvalue
problem from nonlinear spectral mapping}

$\mbox{From}$ now on, we assume that the Hilbert space is discretized so 
that the states are vectors $\psi\in\Cz^K$ and $H$, $W$ are
real symmetric $K\times K$ matrices, $W$ 
diagonal, as, e.g., in the case of a discrete variable 
representation \cite{DVR}.

Let $\<A\>_{\psi}:=\psi^*A\psi/ \psi^*\psi$ define the expectation
value of the operator $A$. The original Hamiltonian matrix has its
spectrum between $\bar{H}-\Delta H$ and $\bar{H}+\Delta H$. The
quantity $\Delta H$ is called the {\it spectral range} and is defined by the
particular basis set used to represent the Hamiltonian. However, 
to simplify the following equations we assume 
without loss of generality that the Hamiltonian matrix is already shifted 
and scaled so that 
\be\label{ab.scale}|\<H-D_0\>_{\psi}|\leq 1\ee 
for any state
$\psi$, where the diagonal real symmetric operator $D_0$ will be specified
later. (In places where the scale factor $\Delta H$ is relevant, it will be inserted explicitly.)

Such a scaling is implemented routinely in the framework
of the Chebyshev polynomial expansion.
We note that this typically moves the ground state energy 
to the lower edge of the spectrum, $E=-1$, while the energies
of interest, including the dissociation energies, appear at the bottom
of the spectrum, $E\sim -1$, 
as the total spectral range
is usually an order of magnitude larger than the physically
relevant spectral range.

We consider the choice
\be\label{ab.3a}
E-W=D_0+{1\over 2}uD_1+{1\over 2}u^{-1}D_2,
\ee
for some complex parameter $u$ and real matrices $D_0$, $D_1$ and $D_2$, diagonal in the
coordinate representation and
satisfying
\be\label{ab.3c}
D_0(r)=0, ~D_1(r)=1, ~D_2(r)=1 \mbox{~~~for } r\in\Omega.
\ee

This is a generalization of the previous results
where the special cases using
$D_0=0$, $D_2=D_1^{-1}$ (as in ref.~\cite{cheb}) and
$D_0=0$, $D_2=1$ (as in ref.~\cite{res_prl}) were encountered.

\begin{figure}
\vspace{8.2cm}
\includegraphics{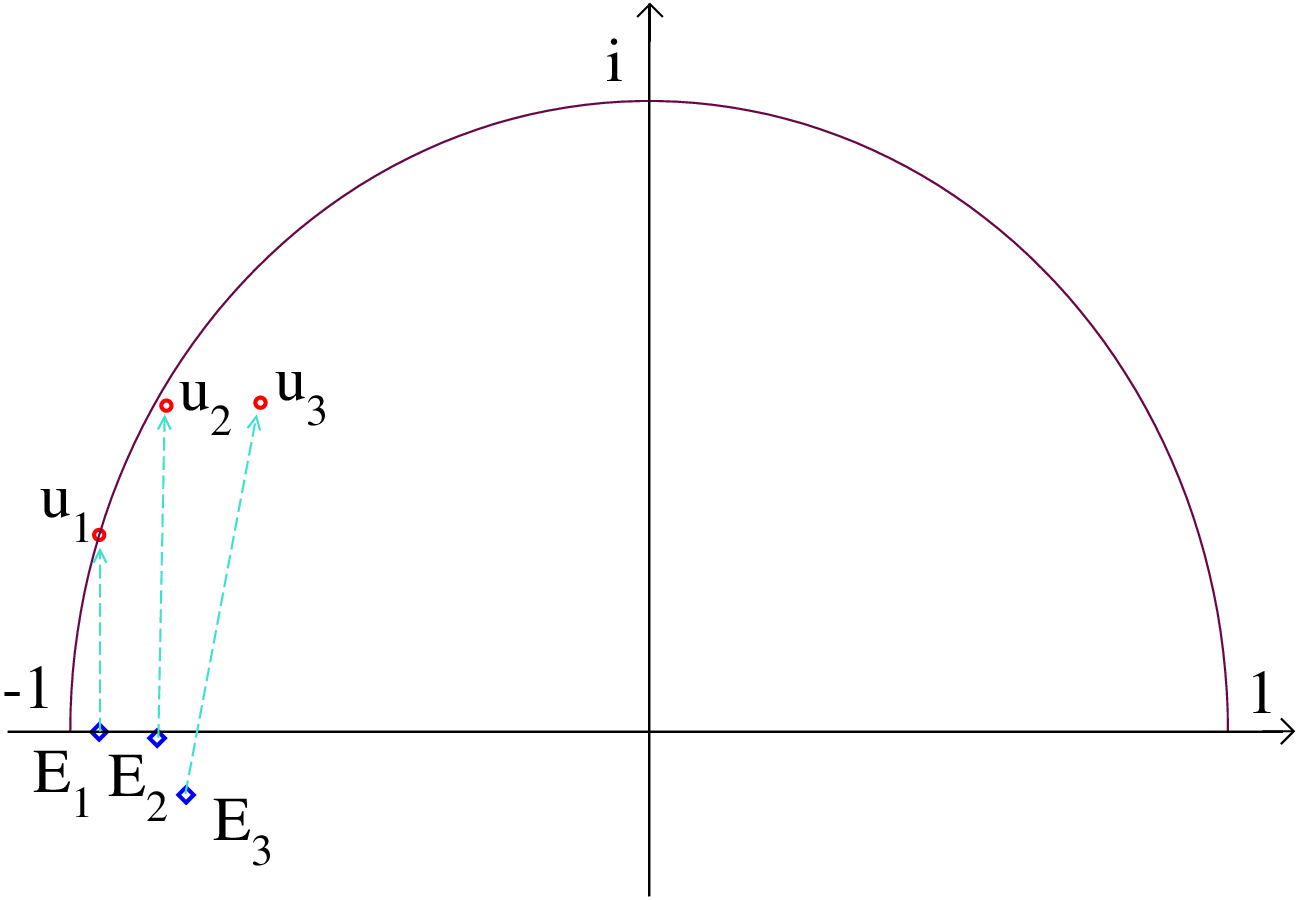}
\caption{\label{fig1} The spectral mapping \Eq{ab.3} maps a bound 
state $E_1$, which is real, to the upper half of the unit circle, 
and the resonance states, $E_2$ and $E_3$, with $\im E_k<0$, inside
the upper half of the unit disk.  $u_2\approx 1$ as it corresponds to a
narrow resonance. (In practice the physically relevant states appear 
in a small subset of the whole
spectral domain.)
}
\end{figure}

To match the original problem in $\Omega$, where the absorbing
potential $W(r)$ vanishes,
$u=u_E$ and $E=E(u)$ must be related by
\be\label{ab.3}
u_E=E+ i\sqrt{1-E^2}, \mbox{~~~or~~~} 
E(u)={1\over 2}u+{1\over 2}u^{-1}
\ee
as shown in \Fig{fig1}.
The absorbing potential then becomes a function of energy $E$
(or, equivalently, of $u=u_E$):
\be\label{ab.WE}
W_E={1\over 2}u(1-D_1)+{1\over 2}u^{-1}(1-D_2)-D_0
\ee 
Insertion of (\ref{ab.3a}) into (\ref{ab.2}) gives a nonlinear 
eigenvalue problem for $u$,
\be\label{ab.4}
   H\psi=\Big(D_0+{1\over 2}uD_1+{1\over 2}u^{-1}D_2\Big)\psi.
\ee
We may think of this equation as an eigenvalue problem 
\be\label{ab.4a}
H\psi=E(u)\psi
\ee
involving an operator-valued $u$-dependent energy 
\be\label{ab.4c}
E(u)=D_0+{1\over 2}uD_1+{1\over 2}u^{-1}D_2
\ee
that, by (\ref{ab.3c}), reduces in $\Omega$ 
to the constant $E={1\over 2}u+{1\over 2}u^{-1}$. 

\section*{Recasting to an ordinary eigenvalue problem}

Apparently, \Eq{ab.4} is equivalent to a linear eigenvalue problem
in a space of doubled dimension:
\be\label{ab.9}
   \hat U\hat\psi=u\hat\psi
\ee
with 2K dimensional state vectors $\hat\psi={\psi\choose\psi'}$ and
square $2K\times 2K$ matrix
\be\label{ab.U}
\hat U:={0 \ \ \ \ \ \ \ \ \ \ \ \ \ \ I \ \ \choose
           -D_1^{-1}D_2  \ \ 2D_1^{-1}(H-D_0)},
\ee
where $I$ denotes the $K\times K$ unit matrix. This can be seen since 
(\ref{ab.9}) yields \[
\psi'=u\psi,~~~D_1^{-1}D_2 \psi-2uD_1^{-1}(H-D_0)\psi+u^2\psi=0.
\]
Now multiplying by $D_1/2u$, we find (\ref{ab.4}). 

We have rewritten the original nonhermitian eigenvalue problem
(\ref{ab.2}) as another nonhermitian eigenvalue problem (\ref{ab.9}),
but with the matrix $\hat U$ of doubled dimension. This would not 
necessarily be
an advantage, if the eigenvalues of $\hat U$ did not possess the 
very important property that
{\bf the spectral domain of $\hat U$ is the unit disk.} 

To see this, we consider an eigenpair $(u,\psi)$ of (\ref{ab.4}) with 
$\psi^*\psi\ne 0$. 
Multiplying (\ref{ab.4}) by $2u\psi^*$ gives the quadratic equation
\[
   u^2\<D_1\>_{\psi}-2u\<H-D_0\>_{\psi}+\<D_2\>_{\psi}=0.
\]

with solutions
\be\label{ab.6}
   u={\<H-D_0\>_{\psi}\pm i\sqrt{\<D_1\>_{\psi}\<D_2\>_{\psi}-\<H-D_0\>_{\psi}^2}\over \<D_1\>_{\psi}}.
\ee

To guarantee correct behavior in the following results we need to
apply the componentwise restrictions on the diagonal damping operators:
\be
\label{ab.cond}
0\leq D_1^{-1}\leq D_2\leq D_1 \ge 1,
\ee 
which together with the condition (\ref{ab.scale}) imply that the square root
is real. Thus, the solutions of (\ref{ab.4}) come in complex 
conjugate pairs, $(u,\psi)$ and $(\bar u,\bar \psi)$. The 
physically relevant eigenenergies with $\im E\leq 0$ come from  
$u$ with $\im u\geq 0$.
(\ref{ab.6}) implies that
\bea\nonumber
|u|^2&=&{\<H-D_0\>_{\psi}^2+\<D_1\>_{\psi}\<D_2\>_{\psi}-\<H-D_0\>_{\psi}^2\over \<D_1\>_{\psi}^2}\\\label{ab.unitdisk}
     &=&{\<D_2\>_{\psi}\over \<D_1\>_{\psi}}\leq 1
\eea
by (\ref{ab.cond}).
Thus, $u$ is a complex number lying in the upper half of the unit disk (see
\Fig{fig1}). Moreover, 
$|u|=1$ if and only if $\<W\>_{\psi}=0$, i.e., if and only if $\psi$ has support in 
$\Omega$, which is the case for the bound states. The states with
$|u|\sim 1$ correspond to the narrow resonances.

\bigskip

The eigenpairs $(u_k,\psi_k)$ of (\ref{ab.4}) can be used to evaluate
the physically interesting quantities (e.g., the complex resonance 
energies $E_k$, scattering amplitudes, etc.). However,
because of the nonlinearity they have somewhat different properties 
from those of the regular nonhermitian eigenvalue problem, which we 
now proceed to derive.

\section*{Completeness}

As was already established, 
the eigenpairs $(u_k,\hat\psi_k)$ of $\hat U$ satisfy
\be\label{ab.11}
   \hat\psi_k={\psi_k\choose u_k\psi_k},
\ee
where $(u_k,\psi_k)$ is an eigenpair of (\ref{ab.4}). Since conversely, 
any such eigenpair determines an eigenpair of $\hat U$,
the nonlinear eigenvalue problem (\ref{ab.4}) has at most $2K$ distinct
eigenvalues. If there are $2K$ distinct eigenvalues $u_1,\dots,u_{2K}$,
the matrix $\hat U$ is diagonalizable, and there is a basis 
$\hat\psi_1,\dots,\hat\psi_{2K}$ of eigenvectors of the form 
(\ref{ab.11}). Therefore, we can obtain the {\it completeness relations},
i.e., for any vector $\hat\phi$
we may write
\be\label{ab.expan2}
   \hat\phi :={\phi_0 \choose \phi_1}=\sum_{k=1}^{2K}\theta_{k}\hat\psi_k=
   {\sum\theta_{k}\psi_k\choose\sum\theta_{k}u_k\psi_k}
\ee
with uniquely determined coefficients $\theta_{k}$.

\section*{Orthogonality}

Using (\ref{ab.4}), the symmetry of $D_0$, $D_1$, $D_2$ and $H$, and
the fact that $D_1$ commutes with $D_2$, we
may compute $\psi_j^{\rm T}H\psi_k$ in two different ways:
\bea\nonumber
\psi_j^{\rm T}H\psi_k
=\psi_j^{\rm T}\left(D_0+{1\over 2}u_kD_1+{1\over 2}u_k^{-1}D_2\right) \psi_k
\\\nonumber =\left(H\psi_j\right)^{\rm T}\psi_k=\psi_j^{\rm T}\left(D_0+{1\over 2}u_jD_1+{1\over 2}u_j^{-1}D_2\right) \psi_k.
\eea
For $j\ne k$, assuming that the eigenvalues are not
degenerate, we may take the
difference, multiply by the factor $2u_ju_k/(u_k-u_j)$, and find 
that the eigenfunctions satisfy the {\it orthogonality 
relations}
\be\label{ab.orth1}
\psi_j^{\rm T}\left(D_2-u_ju_kD_1\right)\psi_k = 0.
\ee
Provided that the left hand side of
(\ref{ab.orth1}) does not vanish for $j=k$, we may normalize the
eigenvectors so that 
\be\label{ab.orth}
\psi_j^{\rm T}\left(D_2-u_ju_kD_1\right)\psi_k = \delta_{jk}.
\ee
Due to (\ref{ab.11}), the orthogonality relations 
can be rewritten in the double-dimension form:
\be\label{ab.orth2}
\hat\psi_j^{\rm T}\hat D\hat\psi_k=\delta_{jk}\ \ \mbox{with}\ \ 
\hat D:={D_2\ \ \ \ \ 0 \ \ \choose  \ 0  \ \ -D_1}.
\ee
Using (\ref{ab.orth2}) and 
(\ref{ab.expan2}) 
we find:
\be\label{ab.coef2}
\theta_{k}= {\sum}_j \theta_{j}\delta_{jk}
={\sum}_j \theta_{j}\hat\psi_j^{\rm T}\hat D\hat\psi_k
=\hat\phi^{\rm T}\hat D\hat\psi_k
\ee

\section*{Resolution of identity}

The orthogonality relations (\ref{ab.orth2}) and completeness imply
the {\it resolution of identity},
\be\label{ab.res1}
\hat I=\sum_{k=1}^{2K}\hat\psi_k\hat\psi_k^{\rm T}\hat D
= \hat D \sum_{k=1}^{2K}\hat\psi_k\hat\psi_k^{\rm T},
\ee
which in the explicit form reads
\bea\label{ab.res2}
&& \sum_{k=1}^{2K}\psi_k\psi_k^{\rm T} =D_2^{-1},\\\nonumber 
&& \sum_{k=1}^{2K}u_k\psi_k\psi_k^{\rm T}=0,\\\nonumber 
&& \sum_{k=1}^{2K}u_k^2\psi_k\psi_k^{\rm T} = -D_1^{-1}.
\eea

\section*{The Green's function}

For $u_E=E+i\sqrt{1-E^2}$ and an eigenpair $(u_k,\psi_k)$ of
(\ref{ab.4}) 
we can write
\[
(H-E+W_E)\psi_k 
={u_E-u_k\over 2 u_ku_E}(D_2-u_ku_ED_1)\psi_k.
\]
Now multiplying from the right by $\frac{2u_ku_E}{u_E-u_k}\psi_k^{\rm
T}$, summing over
$k$ and using the relations (\ref{ab.res2}) we obtain
\bea\nonumber
&&(E-H-W_E)\sum_{k=1}^{2K}\frac{2u_ku_E}{u_E-u_k}\psi_k \psi_k^{\rm
T}\\\nonumber &&\ \ \ \  \
=\sum_{k=1}^{2K}(D_2-u_ku_ED_1)\psi_k\psi_k^{\rm T}=I.
\eea
This leads to a spectral representation of the damped 
Green's function $G_W(E):=(E-H-W_E)^{-1}$
\be\label{ab.GW1}
G_W(E)=\sum_{k=1}^{2K} 
\frac{2u_ku_E}{u_E-u_k}\psi_k\psi_k^{\rm T}.
\ee
This expression uses the eigenvectors, which may be difficult to
generate in large-scale computations.
However, we shall see that (\ref{ab.GW1}) leads to 
iterative schemes for computing matrix
elements of $G_W(E)$. To derive these, we introduce the
double-dimension Green's function 
\bea\label{ab.GW2}
\hat G_W(E) &:=& \sum_{k=1}^{2K} \frac{2u_ku_E}{u_E-u_k}\hat \psi_k
\hat \psi_k^{\rm T} =
{G_W(E) \ * \choose
 \ \ \ * \ \ \ \ \ \ *}
\eea
whose top left corner suffices to represent the
matrix element of the Green's function. Indeed, for suitable 
initial states $\phi_\beta$ and final states 
$\phi_{\alpha}$, we have
\be
\phi_{\alpha}^{\rm T}G_W(E)\phi_{\beta}=\hat \phi_{\alpha}^{\rm T}\hat
G_W(E)\hat \phi_{\beta}
\ee
with 
\be\label{ab.init}
\hat \phi_{\beta}:={\phi_{\beta}\choose 0},\ \ 
\hat \phi_{\alpha}:={\phi_{\alpha}\choose 0}.
\ee

$\hat G_W(E)$ is representable in terms of $\hat
U$ as
\be\label{ab.GW3}
\hat G_W(E)=
(1-\hat U/u_E)^{-1} \hat U\hat D^{-1},
\ee
which can be verified by applying (\ref{ab.GW2}) and
(\ref{ab.GW3}) to $\hat D\hat\psi_k$.
(Note that the term $\hat D^{-1}$ in (\ref{ab.GW3}) is
unnecessary as one usually considers initial states with support
in $\Omega$, where $\hat D^{-1}=1$, but we still prefer to keep
it for completeness.)

\section*{The pseudo-time Schr\"odinger equation}

By replacing $(1-\hat U/u_E)^{-1}$ in (\ref{ab.GW3}) with the geometric
series we can express
$\hat G_W(E)$ as a power series in $\hat U$:
\be\label{ab.power}
\hat G_W(E)=
\sum_{t=1}^\infty u_E^{t-1} \hat U^t\hat D^{-1}.
\ee
This form using a discrete Fourier
transform of the pseudo-evolution operator $\hat U^t$ 
is reminiscent of the integral Fourier transform of the true evolution
operator  
\[
G(E)=i\int_0^\infty e^{-iHt}e^{iEt} dt.
\] 
However, for a given absorbing potential and basis set, $e^{-iHt}$ can
only be represented approximately, while
\Eq{ab.power} is {\it exact} as there is no approximation involved in
evaluating $\hat U^t$. Because $\hat U$ is bounded inside the unit disk
(cf. \Eq{ab.unitdisk}), the expansion (\ref{ab.power}) also 
leads to a numerically stable procedure to compute the Green's function
matrix elements
\be\label{ab.power1}
\phi_{\alpha}^{\rm T}G_W(E)\phi_{\beta}=\sum_{t=1}^\infty 
u_E^{t-1}y_{\alpha\beta}(t),
\ee
where we have introduced the {\it pseudo-time correlation function} 
\bea\label{ab.14}
   y_{\alpha\beta}(t)&:=&\phi_{\alpha}^{\rm T}\phi_{\beta}(t)
=\hat\phi_{\alpha}^{\rm T}\hat\phi_{\beta}(t)
\\\nonumber &=&\hat \phi_{\alpha}^{\rm T}
\hat U^t\hat D^{-1}\hat\phi_{\beta}\ \ (t=0,1,\dots).
\eea

Eqs.~\ref{ab.power1} and \ref{ab.14} constitute an 
important result as they provide a
convenient and efficient numerical framework based on solving
the {\it pseudo-time Schr\"odinger equation}
\be\label{ab.ptse}
\hat\phi_{\beta}(t)=\hat U^t\hat D^{-1}\hat\phi_{\beta}\ \ (t=0,1\dots).
\ee
That is, the quantum dynamics 
problem is finally reduced to the Fourier spectral
analysis of pseudo-time correlation functions.

By noticing that the state vectors $\hat\phi_\beta(t)$ have the
form
\[\hat\phi_{\beta}(t):={\phi_{\beta}(t)\choose \phi_{\beta}(t+1)}\
\ 
\mbox{with}\ \ \hat \phi_{\beta}(0)=\hat D^{-1}{\phi_{\beta}\choose 0},
\]
(\ref{ab.ptse}) can be rewritten as a variant of a damped Chebyshev
recursion formula
\bea\label{ab.12}
   \phi_{\beta}(t)=D_1^{-1}\left[2(H-D_0)\phi_\beta(t-1)-D_2\phi_{\beta}(t-2)\right],
\\\nonumber t=2,3,\dots
\eea
with initial conditions 
$\phi_{\beta}(0)=D_2^{-1}\phi_{\beta}$ and $\phi_{\beta}(1)=0$.

The numerical cost of implementing a single step in (\ref{ab.12}) is
essentially equal to that of multiplication of a real symmetric
$K\times K$ matrix $H$ by a real vector, as the other matrices are diagonal.

\section*{Time doubling}

We note that $\hat D\hat U^t$ is a complex symmetric matrix, 
\[
(\hat D\hat U^t)^{\rm T}=(\hat U^t)^{\rm T}\hat D=\hat D\hat U^t,
\]
which follows from the
spectral representation  
\be\label{ab.Ut}
\hat U^t=\sum_{k=1}^{2K}u_k^t\hat\psi_k\hat\psi_k^{\rm T}\hat D
\ee
or can be verified directly by
the matrix multiplication. 
Thus, for any $t$ and $s$ 
we can write
\bea\nonumber
y_{\alpha\beta}(t+s) &=& \hat\phi_{\alpha}^{\rm T} \hat
U^{t+s} \hat D^{-1}\hat\phi_{\beta}
=\hat\phi_{\alpha}^{\rm T} \hat D^{-1} \hat D\hat U^{s}\hat
U^{t} \hat D^{-1} \hat\phi_{\beta} 
\\\nonumber &=& \left[\hat U^{s}\hat D^{-1}
\hat\phi_{\alpha}\right]^{\rm T} \hat D \hat
U^{t} \hat D^{-1} \hat\phi_{\beta} = \hat\phi_{\alpha}^{\rm T}(s)
\hat D\hat\phi_{\beta}(t),\\\label{ab.yt+s}
\eea
where $\hat\phi_{\alpha}(s)$ and $\hat\phi_{\beta}(t)$ are the
solutions of \Eq{ab.ptse} with initial conditions
$\hat\phi_{\alpha}(0)= \hat D^{-1}\hat\phi_{\alpha}$ and
$\hat\phi_{\beta}(0)= \hat D^{-1}\hat\phi_{\beta}$. 
This means that $y_{\alpha\beta}(t)$ can be computed up to time $2T$ using
\be\label{ab.y2t} 
y_{\alpha\beta}(2t+p)=\hat\phi_{\alpha}(t)^{\rm T}\hat
D\hat\phi_{\beta}(t+p),\ \ p=0,1,
\ee
concurrently with the computation of $\hat\phi_{\alpha}(t)$ and
$\hat\phi_{\beta}(t)$ for
$t=0,\dots,T$. In particular, the calculation of an autocorrelation function
$y_{\alpha\alpha}(t)$ up to time $2T$ requires 
$\sim T$ multiplications of the real and sparse $K\times K$
$H$-matrix on a vector with just a few vectors stored at a time.

\section*{Resonance calculation by harmonic inversion of
pseudo-time correlation functions}

The spectral mapping  $E_k\to u_k$ (\ref{ab.3}) moves all the 
physically relevant eigenvalues
to the vicinity of the unit circle, i.e., they all become extreme
(or nearly extreme for the resonances) eigenvalues of $\hat U$. This,
in turn, creates very favorable conditions for computing these
eigenvalues iteratively using any suitable Krylov subspace method
with the Krylov vectors generated by the powers of $\hat U$, because
the extreme eigenstates contribute most to the latter.
In the present framework, such a strategy of computing the bound and
resonance state energies, in its most simple and
convenient form, boils down to the 
{\it harmonic inversion} of pseudo-time correlation
functions (\ref{ab.14}), which due to (\ref{ab.Ut}) satisfy
\be\label{ab.15}
   y_{\alpha \beta}(t)=\sum_{k=1}^{2K}d_{\alpha \beta k} u_k^t
\ee 
with 
\[
d_{\alpha \beta k}=b_{\alpha k}b_{\beta k}=
\left(\hat\phi_{\alpha}^{\rm T}\hat\psi_k\right)\left(\hat\psi_k^{\rm T}\hat\phi_{\beta}\right).
\]
(Note, that the resolution of identity (\ref{ab.res1}) implies
$\hat\phi_{\alpha}=\hat D\sum_k\hat\psi_k\hat\psi_k^{\rm T}
\hat\phi_{\alpha} = \hat D\sum_k b_{\alpha k} \hat\psi_k$.)

Thus the nonlinear 
eigenvalue problem (\ref{ab.4}) is reduced to the signal
processing problem of 
finding the spectral parameters
$(u_k,d_{\alpha\beta k})$ ($k=1,\dots,2K$) satisfying (\ref{ab.15}) for the 
sequence(s) $y_{\alpha\beta}(t)$ computed by (\ref{ab.12}) and (\ref{ab.14}). 
The simplest procedure then corresponds to considering a single
doubled autocorrelation function $y_{\alpha\alpha}(t)$ with $t=0,\dots,2T$. 
In exact arithmetic the harmonic inversion of such a sequence 
will give the exact results if $T> 2K$, thus, using only 
$T\sim 2K$ of real matrix-vector products. 
However, this is impractical as 
it would formally require to solve a $T\times T$ eigenvalue problem.
To reduce the computational burden and to maintain numerical 
stability the eigenvalues are extracted very efficiently in a small 
Fourier subspace by the FDM 
\cite{Neuh95,filter2_JCP}. If we introduce the unscaled energy
$\varepsilon=E\Delta H+\bar{H}$,
the required length $2T$ of the 
doubled sequence needed to converge an eigenenergy $E_k$ (cf. \Eq{ab.3})
by the FDM will be defined by the locally averaged density of states
$\rho(\varepsilon)$ for $E_k\sim E$ and the spectral range $\Delta H$
of the Hamiltonian matrix
according to the approximate relationship \cite{filter2_JCP,ManCarr}
\be\label{ab.conv}
T \ge 2\pi\Delta H\rho(\varepsilon)\sqrt{1-E^2},
\ee
where the factor $\sqrt{1-E^2}$ arises from the $u\to E$ mapping
(\ref{ab.3}). 

Because the eigenpairs come in complex 
conjugate pairs $(u_k,\psi_k)$ and $(\bar u_k,\bar \psi_k)$, once the
initial vectors $\hat\phi_\alpha$ and  $\hat\phi_\beta$ are real
(\ref{ab.15}) becomes
\be\label{ab.real}
   y_{\alpha \beta}(t)=\re\sum_{k=1}^{K}d_{\alpha \beta k} u_k^t,
\ee 
where only the physical eigenvalues with $\im u_k > 0$ are included in
the sum.
Note also that
for the pure bound state problem, when $|u_k|=1$ and the
eigenfunctions $\psi_k$ 
(and, therefore, the coefficients $d_{\alpha \beta k}$) are real, the
sequence $y_{\alpha \beta}(t)$ has the time reversal symmetry
\be\label{sym}
   y_{\alpha \beta}(-t)=\bar{y}_{\alpha \beta}(t)
\ee 
which further doubles the total time by extending the signal to
the negative times.

\section*{Inversion of the time cross-correlation functions}

As was previously shown in ref.~\cite{cross} one can, in principle, compute
matrix elements $\phi_\alpha^{\rm T} G_W(E) \phi_\beta$ for {\it any}
$\phi_\alpha$, $\phi_\beta$ by (i) propagating a {\it single} initial
state $\phi_0$ using (\ref{ab.12}), (ii) computing the cross-correlation functions
$y_{00}(t)$, $y_{0\alpha}(t)$ and $y_{0\beta}(t)$ using (\ref{ab.14})
and (iii) solving the corresponding harmonic inversion problems
(\ref{ab.15}) for the unknown parameters $\{u_k,b_{\alpha k},b_{\beta
k}\}$ to evaluate  
\be\label{ab.Gcross}
\phi_{\alpha}^{\rm T}G_W(E)\phi_{\beta}=\sum_{k=1}^{2K} 
\frac{2u_ku_E}{u_E-u_k} b_{\alpha k}b_{\beta k}.
\ee
Moreover, a more stable evaluation of the matrix element
$\phi_{\alpha}^{\rm T}G_W(E)\phi_{\beta}$
may be
achieved using the {\it Regularized Resolvent Transform} (RRT) which
was described in detail in ref.~\cite{RRT_JPC}. The advantage of RRT
is that the calculation of spectral parameters $\{u_k,b_{\alpha
k},b_{\beta k}\}$ is avoided, the spectra are computed directly 
by matrix inversion or by solving linear systems.

This results in an enormous numerical saving as, traditionally, the
time-dependent approaches require multiple initial state propagations
in order to compute, for example, the full S-matrix or cumulative
reaction probability.
However, the described approach, although formally exact (in exact
arithmetic), is applicable only when the dynamics is governed solely by
narrow resonances: the parameters of 
very broad resonances (or poles of the Green's function)
are generally grossly inaccurate, leading to
very unstable spectral estimation by \Eq{ab.Gcross}. A significant
numerical saving in cases with broad resonances 
is, however, achievable if one adapts
another strategy in which one (i) propagates a set of initial states
$\{\phi_\beta\},\ (\beta=1,\dots,L)$ using \Eq{ab.12}, (ii) computes the
{\it cross-correlation matrix} using the doubling formula
(\ref{ab.y2t}) and (iii) solves the harmonic inversion problem (\ref{ab.15}) for the 
$L\times L\times 2T$ data matrix (see
refs.~\cite{Neuh95,cross,RRT_JPC}). As argued in ref.~\cite{cross},
this approach has a potential of reducing the total propagation time
$T$ required for the accurate harmonic inversion by a factor of $L$,
i.e., preserving the total number of matrix-vector
multiplications, $N_{\rm total}=L\times T$. 
Unfortunately, to use the doubling trick (\ref{ab.y2t}) in this case, 
the storage requirement to generate the
cross-correlation matrix is increased because of the
need to simultaneously propagate $L$ states rather than one.

\section*{Practical considerations: choosing the absorbing potential}

The physical observables can generally be computed from the Green's 
function matrix
elements at real energies. For this reason and for the sake of
simplicity in the following analysis we will assume
the energy $E$ to be real. However, essentially the same conclusions 
hold for complex energies near the real axis, i.e.,
including the narrow resonances. 

Since $u_E^{-1}=E-i\sqrt{1-E^2}$, our construction (\ref{ab.WE}) leads 
to the complex valued absorbing potential
\bea\nonumber
W_E = E\left(1-{D_1+D_2\over 2}\right)-D_0
- i\sqrt{1-E^2} {D_1-D_2\over 2}\\\label{ab.WE1}.
\eea
The restrictions (\ref{ab.cond}) on the choice of $D_1$ and $D_2$,  
imply that $\im W_E\le 0$. This is essential since it leads to the 
correct limit for the Green's function $G_W(E)$ on the real line. 
The real part $\re W_E$ is generally nonzero; although its role is not 
crucial in scattering calculations, significant values of $\re W_E$ 
may or may not be desirable.
In particular, a positive $\re W_E$ reduces the total density 
of states $\rho(E)$, while affecting only slightly the
resonance states, and thus, it may accelerate the convergence (cf. 
\Eq{ab.conv}). At the
same time $\re W_E$ may get excessively large compared
to $\im W_E$, for example, near the spectral edge, $E\sim -1$,  if
$D_0=0$.  It is not absolutely clear without numerical tests 
how this artifact will affect the
results, while
it can be controlled if the matrix $D_0$, which is in principle
unrestricted, is used.
Namely, relatively to the total spectral range $\Delta H$ (which
becomes unity after scaling of $H$), 
the physical energy range is generally only 
a small part of the unit disk. Therefore, within this range we can
often assume $W_E\approx W_{E_0}$ for some reference energy $E_0$.
Now by setting
\be \label{ab.D0}
D_0=E_0\left(1-{D_1+D_2\over 2}\right),
\ee
we can
significantly reduce (or increase) the real component if needed,
\be \label{ab.reW-D0}
\re W_E = (E_0-E)\left({D_1+D_2\over 2}-1\right),
\ee
without affecting $\im W_E$.
In order to avoid negative $\re W_E$ 
in the physically relevant energy region, it suffices
to take $E_0$ higher than the maximum energy of interest.

Clearly, the behavior of the absorbing potential $W_E$ where it
starts to turn on
has the main effect for the scattering calculations.
In fact, as follows from our numerical tests, the best 
performance is achieved when both $D_1\approx 1$ and
$D_2\approx 1$ over almost the entire grid used to represent the state
vectors. Therefore, it is convenient to use the real operators  
$\gamma_1$ and  $\gamma_2$ defined by
\be
D_1=e^{\gamma_1},\ \ D_2=e^{-\gamma_2},
\ee
which vanish in $\Omega$ and slowly turn on
in the absorbing region, and assume that both $\gamma_1$ and  $\gamma_2$
are small. By expanding into the Taylor series up to the second order, 
we can rewrite (\ref{ab.WE1}) as
\bea \label{ab.WMexp}
W_E &\approx& (E_0-E)\left({\gamma_1-\gamma_2\over 2} +
{\gamma_1^2+\gamma_2^2\over 4}\right)\\\nonumber
&-& i\sqrt{1-E^2} \left({\gamma_1+\gamma_2\over 2}+
{\gamma_1^2-\gamma_2^2\over 4}\right).
\eea

We expect the results to be generally insensitive to a particular form
of $W_E$ as long as $\gamma_1$ and  $\gamma_2$ are sufficiently smooth
and have sufficiently large spatial extension. So here we only consider the two
special choices \cite{cheb,res_prl}

\bigskip
\noindent
{\bf M\&T scheme:}\ \ $\gamma_1\equiv\gamma_2\equiv\gamma$ as in
ref~\cite{cheb} with
$\gamma(R)>0$:
\be\label{ab.WMT}
W_E = (E_0-E)(\cosh(\gamma)-1)
- i\sqrt{1-E^2}\sinh(\gamma).
\ee
As follows from the expansion (\ref{ab.WMexp}), 
$\re W_E$ is already small as
its leading term becomes quadratic in $\gamma$, 
while $\im W_E$ is linear:
\be\label{ab.WMT2}
W_E\approx {1\over 2}(E_0-E)\gamma^2 
-i\sqrt{1-E^2}\gamma.
\ee
Note again that in the computations involving molecular vibrational spectra
the energies of interest are usually close to the bottom of the
spectral range of the
scaled Hamiltonian matrix, $E\sim -1$. Defining the shifted energy
$\varepsilon:=E+1$ with zero at the bottom of the spectrum and
assuming $\varepsilon\ll 1$, we can approximately determine how the absorbing 
potential depends on energy:
\be\label{ab.WMT1}
W_E\approx {1\over 2}(\varepsilon_0-\varepsilon)\gamma^2 
- i \sqrt{2\varepsilon}\gamma.
\ee

A convenient choice for $\gamma$ in \Eq{ab.WMT}  corresponds to 
$\gamma(R)$ being a function of the reaction coordinate $R$, vanishing in
the interaction region, $R<R_{\Omega}$, and smoothly growing
in the absorbing region, $R_{\Omega}<R<R_{\rm max}$, where $R_{\rm max}$
defines the fartherst grid point:
\be \label{ab.gamma} \gamma(R) =
{\lambda\over \sqrt{\Delta H}}
\left({R-R_{\Omega}\over R_{\rm max}-R_{\Omega}}\right)^2.
\ee
Here $\lambda$ is an adjusting
parameter.
As follows from \Eq{ab.WMT1}, the factor $\sqrt{\Delta H}$ in
\Eq{ab.gamma} minimizes the
sensitivity of $\im W_E$ to
the actual spectral range $\Delta H$ of the Hamiltonian matrix. Note
that with this construction $\re W_E$ will be sensitive to $\Delta H$,
but generally small. 

\bigskip

\noindent
{\bf N\&M scheme:}  $\gamma_2\equiv 0$ and $\gamma_1\equiv2\gamma$, 
as in ref~\cite{res_prl}:
\bea\label{ab.WNM}
W_E &=& {E_0-E-i\sqrt{1-E^2}\over 2}\left(e^{2\gamma}-1\right)\\
&\approx& (E_0-E-i\sqrt{1-E^2})\gamma,
\eea
where we retained only the leading linear term. 
Further assuming $\varepsilon$ to be small, we obtain
\be\label{ab.WNM1}
W_E\approx (\varepsilon_0-\varepsilon)\gamma 
- i \sqrt{2\varepsilon}\gamma.
\ee
Thus the difference between M\&T and N\&M schemes is
the different behavior of $\re W_E$. This difference may be eliminated
almost completely by manipulating with $D_0$. 
In the next section, we examine the effect of this difference on the
quality of the resonance calculations.

\bigskip

\begin{figure}
\vspace{12.cm}
\includegraphics{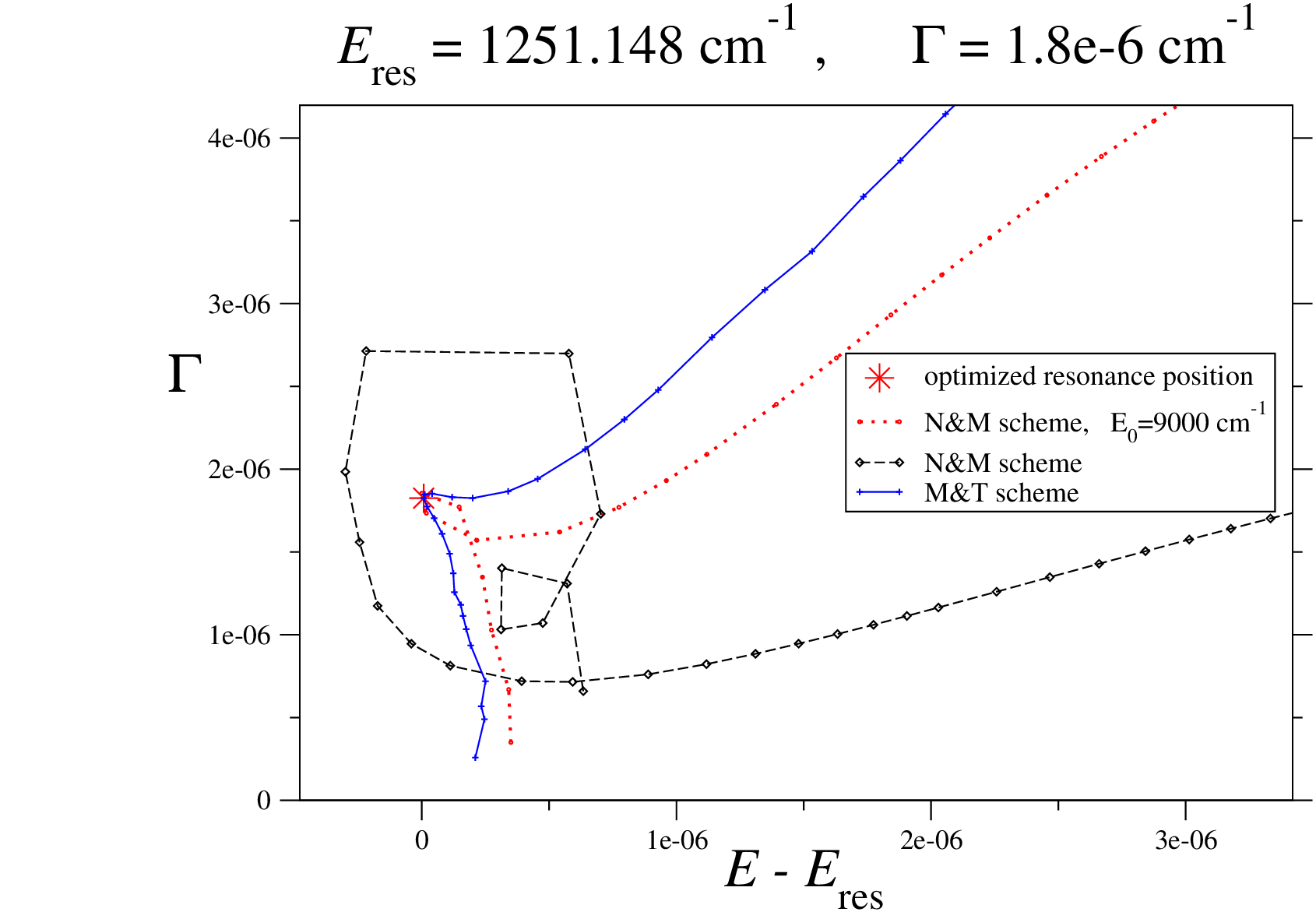}
\includegraphics{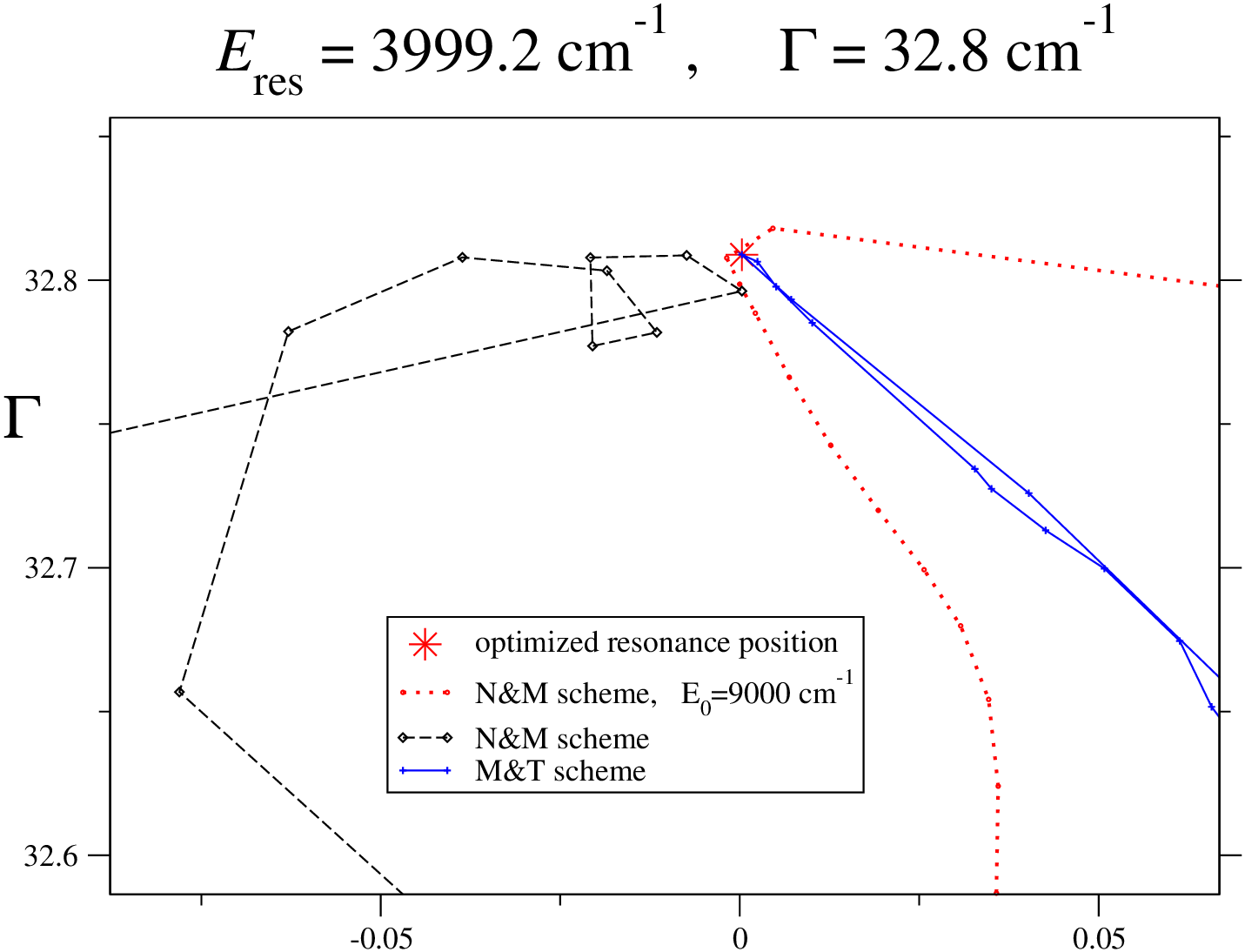}
\caption{\label{fig2}Typical eigenvalue trajectories for two resonance
states obtained using three different damping schemes (see text).
}
\end{figure}

In order to obtain the best resonance energy estimates for given
basis set, the common practice is to search for stationary eigenvalues
$E_k$ under variations of the absorbing potential
\cite{Jolicard,RisM,cusp}. 
This is
usually done by varying its amplitude, which here corresponds
to the free parameter $\lambda$. 
(One could possibly change instead $E_0$, or treat $E_0$ as a function 
of $\lambda$.)
The eigenvalue trajectories $E_k(\lambda)$ are then analyzed for
stationary points, such as the point of maximum curvature, which in an
ideal case may be a cusp (see \Fig{fig2}). 
One way to practically find such a point is to minimize the
derivative  $dE_k/d\lambda$. 
The generation of the eigenvalue trajectories 
usually significantly increases the computational time, while a
reasonable accuracy for most states can be achieved from a single or a
few runs using an {a priori} established optimal value of $\lambda$.
Therefore, such a search is justified only if a very high accuracy 
for all resonances is needed. 
The cases where eigenvalue trajectory analysis may be necessary
include extremely narrow (shape) resonances.

\section*{Numerical Example: Resonances of the HCO molecule.}

In this section we apply the present approach for an accurate computation
of the nonrotating HCO resonances. This system has been used in the 
past by various groups as a
benchmark problem to test new approaches (see the comprehensive review
\cite{poirier} and references therein). 
In particular, the first resonance calculation using FDM was applied
to HCO \cite{HCO}.
We use the potential energy surface of Keller {\it et al}
\cite{keller}, where a
resonance calculation was also performed and the resonances were assigned.
Recently, Poirier \& Carrington  \cite{poirier}
applied the above mentioned PIST method to reproduce these results but 
with a significantly higher accuracy and numerical efficiency. We use
the latter results as a reference.
Although, PIST is a very efficient alternative to the present
technique, in our study we do not make an adequate comparison of 
the two methods as it would require the use of, at least, the same
basis set. 

The present choice of the basis is the most primitive direct-product
Jacoby coordinate DVR grid, similar to that used in ref.~\cite{HCO}. 
Several
basis sets with various grid sizes and cutoff parameters were tested. 
The results reported below correspond to
a particular single set of parameters, which, as was verified
empirically, provided extremely accurate
results for the resonances in the energy region below 9000 cm$^{-1}$.
In order to avoid an extensive search for optimal basis set parameters,
our strategy is to use a larger and denser grid than may be necessary
to garantee high accuracy. This strategy is justified by a very
favorable numerical scaling of the present technique with the basis size.

For $R_{\rm H-CO}$ (the dissociation 
coordinate) we used $160$ 
sinc-DVR \cite{sincDVR} points in the interval $[2,8]$ a$_0$, contracted
to $N_R=40$ points by means of the HEG \cite{HEG} method  
using the eigenfunctions of the 1D Hamiltonian defined by 
$V(R)=\mbox{min}_{r,\theta}V(R_e,r,\theta)$, where $V(R_e,r,\theta)$
is the 3D potential of HCO in the Jacoby coocrdinates. 
For the $r_{\rm CO}$ coordinate 
we used $64$ sinc-DVR points in the interval $[1.8,3.5]$ a$_0$, 
contracted to $N_r=16$ points. For the angular 
variable we used $N_\theta=46$ Gauss-Legendre-quadrature DVR points.
In order to reduce the Hamiltonian spectral range $\Delta H$ we
replaced the high values of the potential by the cutoff value $V_{\rm
cut}=25000$ cm$^{-1}$. This resulted in $\Delta H=44654$ cm$^{-1}$.
We note, that for the present approach the quality of the grid basis is
measured by $\Delta H$, which directly affects the convergence
(cf. \Eq{ab.conv}). Since
our basis is very primitive the spectral range is relatively high.
One may argue though that a primitive basis has the advantage of
minimizing the number of adjusting parameters and leads to a fast matrix-vector
multiplication for given grid size.

Here we report results corresponding to the three different schemes:
(I) N\&M scheme with $D_0$ defined by $E_0=9000$ cm$^{-1}$ (the maximum energy of interest), (II) N\&M scheme with
$D_0=0$, and (III) M\&T scheme with $D_0=0$. (We note that the use of
nonzero $D_0$ in the M\&T scheme has
almost no effect as $\re W_E$ is already very small in this case.)

For each scheme the damping 
potential has been taken in the form (\ref{ab.gamma})
with $R_{\Omega}=5$ a$_0$. For this study we varied the strength
parameter $\lambda$ in the range $[0.0002,0.6]$ with a
logarithmic distribution of 22 $\lambda$ values.

For each $\lambda$ an autocorrelation function $y(t)$
($t=0,...,2T=20000$) with a random initial
vector was generated using $T=10000$ matrix-vector
multiplications. Calculation of a single autocorrelation function
takes 239 sec on an Athlon 1.8 GHz pc using the gnu g77 Fortran compiler.  
Because the latter is very inefficient, the reported timing only 
gives a rough estimate of the algorithm efficiency. 
The autocorrelation functions were then inverted by our
quadruple-precision FDM code \cite{FDM32}. For a large enough value of
$\lambda$, the FDM results converge
for most resonance states using $T\approx 7000$, however, to maintain an
extremely high accuracy, especially for the lowest sharp resonance
$|013)$ with width $\Gamma=3\cdot 10^{-8}$ cm$^{-1}$ and
for small $\lambda$ values, we needed $T\approx 10000$. (The effective
density of states, that includes both the true resonance states and
states representing the continuum, becomes high for small $\lambda$.)
Of course, a smaller grid and lower $V_{\rm cut}$ would result in a
smaller $\Delta H$ reducing the needed number of iterations accordingly.  
Also note that because only small
generalized eigenvalue problems are encountered in FDM, the cpu-time
increase due to the use of the quadruple precision is not essential,
while it improves the accuracy and accelerates the convergence
of the extracted eigenvalues
significantly, compared to the double precision code. 
For the present
``toy problem'' the harmonic inversion part is as time-consuming (with
quadruple precision) as the 
correlation function generation, however, 
the former becomes relatively negligible
for bigger systems.

Let $E_{\rm res}:=\re E_k$ define the positions and $\Gamma_k:=-2\im
E_k$, the widths of the resonances.
The eigenvalue trajectories $E_k(\lambda)$ with $\Gamma_k < 120$ cm$^{-1}$ are plotted in the
complex plane and the most stationary point (see above) is assumed as the best
resonance energy estimate.
In \Fig{fig2}
we show the eigenvalue trajectories generated using the three
different damping schemes decribed above  for two states (a) 
with $E_{\rm res}=3999.2$, $\Gamma=32.8$ cm$^{-1}$ and 
(b) $E_{\rm res}=1251.148$,
$\Gamma=1.8\cdot 10^{-6}$ cm$^{-1}$.
Schemes I and III 
result in quite
different trajectories for most $\lambda$ values but have easily
identified stationary points (cusps) for about the same value
$\lambda$. These stationary points
are very close to each other. At the same time, scheme II gives more 
complex trajectories that circle around the resonance position, but have
no well identified stationary point.
Similar patterns are observed for most other states.
Note also, that for some resonances the cusps
may not exist, however having the two types of
trajectories helps one to better locate the complex resonance energy
by taking the point where the
trajectories of different types approach each other.
Furthermore, scheme III generally results in sharpest cusps, but
because it corresponds to a smaller $\re W_E$, it also requires larger
values of $T$ for an accurate harmonic inversion, as smaller $\re W_E$
corresponds to higher density of states. For the same reason 
scheme II needs the smallest $T$.
\begin{figure}
\vspace{12cm}
\includegraphics{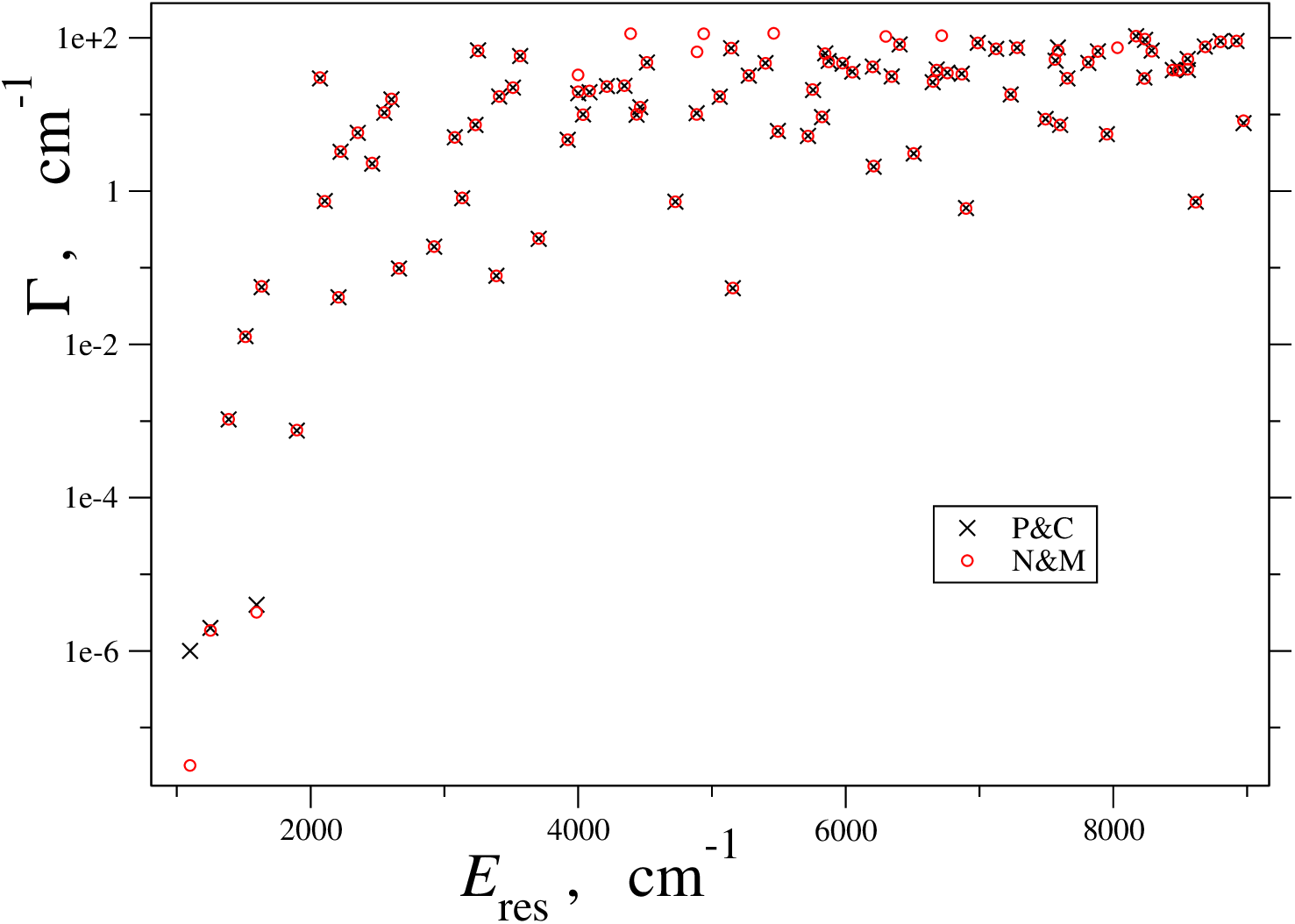}
\includegraphics{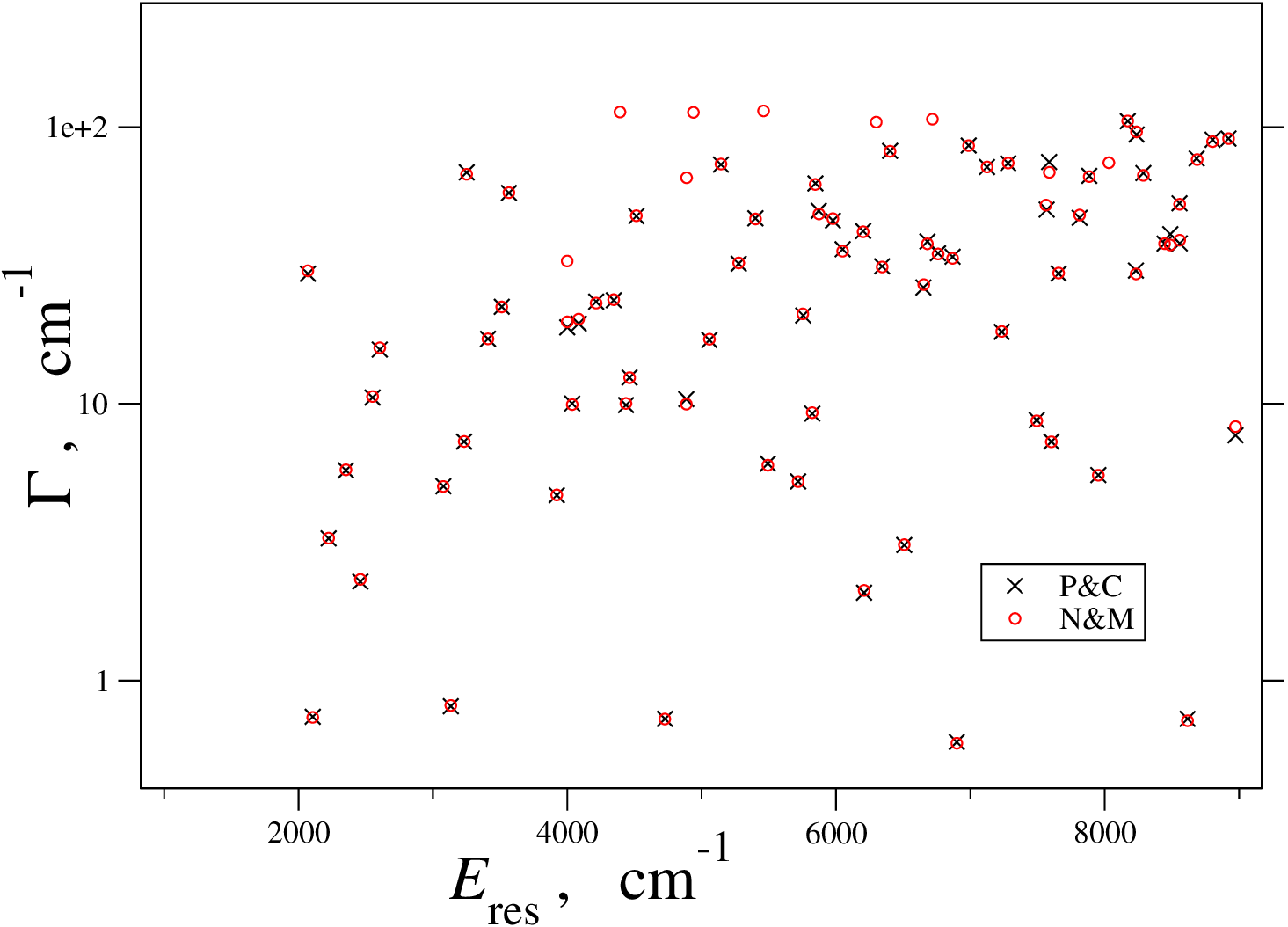}
\caption{\label{fig3} The widths $\Gamma$ versus resonance positions
$E_{\rm res}$ shown for two different scales. The present results (circles) 
are obtained using scheme I (see text). The crosses are the results of
Poirier \& Carrington \cite{poirier}.
}
\end{figure}

Surprisingly, for most resonances, the stationary point in the
eigenvalue trajectories are approached for the same value
$\lambda_{\rm opt}=0.08$, while only 
for a few extremely sharp resonances near the
dissociation threshold $E_{\rm dis}=1086$ cm$^{-1}$, the values 
$E_k(\lambda=0.08)$ give poor estimates of the true complex resonance
energies. For these states the optimal $\lambda$ is a monotonically
growing function of $E_{\rm res}-E_{\rm dis}$; 
the optimal $\lambda$ for the sharpest state, $|013)$,
is $\lambda_{\rm opt}\approx 0.0016$ and for the next state, $|005)$,
it is  $\lambda_{\rm opt}\approx 0.0064$.  By setting $\lambda=0.08$ 
with the present choice of $\gamma(R)$, one can use
a single autocorrelation function to obtain good resonance
estimates for most states, but a few.

There is no simple way to
determine the error bounds other than by comparing the results using
different parameters. The accuracy in the positions
of the resonances varies depending
mostly on $\Gamma_k$. It also depends in a less systematic way on
the quantum numbers.  For narrow resonances ($\Gamma_k\sim 1$ cm$^{-1}$
or less) the accuracy in the
position is of the order of $0.01$ cm$^{-1}$ or better, while for the
other states with $\Gamma_k < 100$ cm$^{-1}$ it is generally of the
order of $0.1$ cm$^{-1}$. We believe, that generally the error in the
reported widths is less than 1\%, although this may not be true for
some states.

\Fig{fig3} and
Table I summarize our results obtained with scheme I 
and compare them with the calculation of
Poirier \& Carrington \cite{poirier}. 
The assignments of the resonance states have
been made by Keller {et al} \cite{keller}. 
The latter results for the resonance
parameters are not displayed here as they are less accurate.
Our results generally agree very well with the results of Poirier \& Carrington,
although some states are missing in the latter, such as the broad
resonance shown in \Fig{fig2}: missing eigenvalues are a common
drawback of Lanczos based diagonalization approaches. 
Note also that for the first two sharp resonances, 
Poirier \& Carrington did not report reliable width estimates.

Even though under the present circumstances a fair comparison between
the present approach and PIST \cite{poirier} is not possible (e.g., very
different basis sets were used in the two cases), the
following remarks seem appropriate.
At least for the HCO case, PIST appeared very
efficient (required relatively few matrix-vector multiplications to
accurately compute the resonances) and relatively small cpu-time,
although the fact that it has more adjusting
parameters than the present technique may be viewed as its
disadvantage.
To conclude, the present approach, at least, for the resonance
calculations, appears very reliable, accurate and efficient. It requires a
minimal number of adjusting parameters and scales favorably with the
size of the system. The method is quite flexible in the
choice of the damping scheme, but according to our tests
an optimal choice corresponds to scheme I (see above) with
$\lambda\approx 0.08$.

We make the corresponding Fortran codes available upon
request (e-mail: mandelsh@uci.edu).

\bigskip
\noindent
{\bf Acknowledgement.}\hskip 0.2in We are grateful to Isabella Baccarelli for
careful reading of the manuscript and Hua Guo, Bill Poirier and Tucker
Carrington for useful discuissions. We also thank Reinhard Schinke for
sending us the HCO potential.
V.A.M. acknowledges the NSF support, grant CHE-0108823. V.A.M. is
Alfred P. Sloan research fellow.



\newpage

\begin{table}[h]
\caption{Resonance energies $E_{\rm res}$ and widths $\Gamma$
(cm$^{-1}$) for the nonrotating HCO molecule. 
The quantum numbers describe, respectively,
the CH stretch, the CO stretch, and the bend. (The last digit in
the ``present results'' gives a rough estimate of the computational error, although
an accurate error estimate is generally unknown.)}
\begin{tabular}{ccccc|ccccc} \\ \hline
\multicolumn{1}{c}{\underline{Keller {\it et al} \cite{keller}}} 
& \multicolumn{2}{c}{\underline{Poirier and Carrington \cite{poirier}}} 
& \multicolumn{2}{c}{\underline{\ \ \ Present results\ \ \ }} &
\multicolumn{1}{c}{\ \underline{Ref. \cite{keller}}} 
& \multicolumn{2}{c}{\underline{Poirier and Carrington \cite{poirier} }} 
& \multicolumn{2}{c}{\underline{\ \ \ \ Present results\ \ \ \ }} \\
State & $E_{\rm res}$ & $\Gamma$ & $E_{\rm res}$ & $\Gamma$ & State & $E_{\rm res}$ & $\Gamma$ & $E_{\rm res}$ & $\Gamma$ \\
\hline
$|013)$	&	1098.7934& $\le$1e-6 &  1098.7963 &    3e-8 & $|042)$	&	5492.86	& 6.08 	&	5492.91  &      6.00 \\
$|005)$	&	1251.148&  $\le$2e-6 &   1251.1525 &    1.8e-6 & $|034)$	&	5717.85	& 5.245 &	5717.82  &      5.23 \\
$|111)$	&	1386.880& 1.05e-3 &	1386.8770 &     1.05e-3 &	&	5756	& 20.85 &	5752.56  &      21.13 \\
$|103)$	&	1512.993& 0.0128&	1512.986  &     0.0126 &$|140)$	&	5823.12	& 9.22  &	5823.02  &      9.29 \\
$|030)$	&	1595.972& 4e-6& 	1595.9708 &     3.2e-6 & $|026)$	&	5846.5	& 62.6 	&	5845.4    &     62.0 \\
$|201)$	&	1633.526& 0.0558&	1633.518 &      0.0572 &	&	5872	& 49.8 	&	5872.7    &     48.5 \\
$|022)$	&	1897.242& 7.5e-4 &	1897.247 &      7.60e-4 &$|230)$	&	5976.83	& 45.9  &	5976.3   &      46.7 \\
$|300)$	&	2069.6	& 29.5 	&	2069.2 &        30.3 &$|132)$	&	6049.7	& 36.2  &	6049.3   &      35.6 \\
$|014)$	&	2105.537& 0.741 &	2105.536 &      0.737 &	&	6201.9	& 42.1  &	6201.8   &      41.8 \\
$|006)$	&	2208.16	& 0.0412 &	2208.198 &      0.0411 &$|051)$	&	6209.04	& 2.079 &	6209.07  &      2.12 \\
$|120)$	&	2223.85	& 3.265 &	2223.83 &       3.274 &	&	   	&	&	6299.7   &      104.2 \\
$|202)$	&	2352.47	& 5.75 	&	2352.38 &       5.77 &$|320)$	&	6343.3	& 31.4	&	6343.9   &      31.2 \\
$|112)$	&	2460.12	& 2.28 	&	2460.11 &       2.32 &$|222)$	&	6402.7	& 82.01 &	6402.3   &      81.7 \\
$|210)$	&	2550.78	& 10.54 &	2550.71  &      10.63 &$|043)$	&	6507.64	& 3.09 	&	6507.59  &      3.10 \\
$|104)$	&	2604.30	& 15.7 	&	2604.40  &      15.93 &$|019)$	&	6650.3	& 26.32 &	6652.99  &      26.96 \\
$|031)$	&	2660.63	& 0.09753 &	2660.640 &      0.0981 &$|035)$	&	6680	& 38.6 	&	6679.8   &      37.9 \\
$|023)$	&	2923.556& 0.1881 &	2923.554 &      0.1888 &	&	   	&	&	6717.6   &      106.8 \\
$|015)$	&	3077.5	& 5.04 	&	3077.53  &      5.03 &$|027)$	&	6759.11	& 35 	&	6759.6   &      34.8 \\
$|007)$	&	3132.82	& 0.8077 &	3132.991 &      0.814 &$|141)$	&	6867.6	& 34 	&	6867.6   &      33.5 \\
$|121)$	&	3232.32	& 7.3 	&	3232.327 &      7.31 &$|060)$	&	6899.15	& 0.6 	&	6899.107 &      0.594 \\
$|203)$	&	3251.1	& 68.6	&	3251.3   &      67.5 &$|223)$	&	6987	& 85.8 	&	6986.8   &      85.7 \\
$|040)$	&	3388.21	& 0.0786 &	3388.200 &      0.0782 &$|125)$	&	7124.2	& 71.7 	&	7124.2    &     71.7 \\
$|113)$	&	3409.6	& 17.17 &	3409.66  &      17.18 &$|052)$	&	7233.1	& 18.2  &	7233.23  &      18.26 \\
$|211)$	&	3512.31	& 22.43 &	3512.29  &      22.39 &	&	7281	& 74  	&	7281.1   &      74.1 \\
$|105)$	&	3566.1	& 57.8 	&	3565.61  &      57.9 &$|044)$	&	7493.83	& 8.73 	&	7493.84  &      8.68 \\
$|032)$	&	3704.872 & 0.238 &	3704.866 &      0.240 &	&	7568	& 50.3  &	7563.4   &      52.2 \\
$|024)$	&	3921.77	& 4.677 &	3921.77   &     4.687 &	&	7585.8	& 74.8  &	7587.2   &      68.6 \\
$|130)$	&	3999.9	& 18.9	&	3999.84  &      19.80 &$|150)$	&	7602.67	& 7.31	&	7602.63  &      7.29 \\
	&	   	&	&	3999.2   &      32.8 &$|036)$	&	7656.7	& 29.56	& 	7656.64  &      29.68 \\
$|016)$	&	4036.49	& 10.03 & 	4036.63  &      9.94 &$|142)$	&	7813.3	& 47  	&	7813.1   &      48.1 \\
$|220)$	&	4084.6	& 19.5  &	4084.71  &      20.25 &$|240)$	&	7884.9	& 66.6 	&	7884.9   &      66.2 \\
$|122)$	&	4214.7	& 23.4 	&	4214.49  &      23.12 &$|061)$	&	7952.3	& 5.53 	&	7952.29   &     5.52 \\
$|114)$	&	4345.8	& 23.68	&	4345.79  &      23.80 &	&	   	&	&	8030.7    &     74.3 \\
	&	   	&	&	4392.1   &      113.4 &$|224)$	&	8171.1	& 105	&    	8170.6   &      105.1 \\
$|041)$	&	4436.7	& 9.9	&	4436.61  &      10.03 &$|151)$	&	8232	& 30.3 	&	8233.06  &      29.48 \\
	&	4463.8	& 12.45 &	4463.70  &      12.44 &$|053)$	&	8237.3	& 94   	&	8236.6   &      96.1 \\
$|212)$	&	4514.2	& 47.65 &	4514.1   &      47.8 &	&	8287.1	& 68.2  &	8287.2   &      66.9 \\
$|033)$	&	4726.773& 0.728 &	4725.768 &      0.727 &$|045)$	&	8445.08	& 37.92	& 	8444.9   &      37.9 \\
$|017)$	&	4885.6	& 10.4 	&	4885.78  &      9.98 &	&	8488	& 41 	& 	8486.7   &      37.5 \\
	&	   	&	&	4887.8   &      65.6 &	&	8556.2	& 53	& 	8557.5   &      52.6 \\
	&	   	&	&	4938.4   &      112.9 &$|151)$	&	8558	& 38 	&	8557.7   &      39.0 \\
$|131)$	&	5057.9	& 17 	&	5057.81  &      17.12 &$|070)$	&	8616.5	& 0.728	& 	8616.421 &      0.717 \\
$|213)$	&	5143.5	& 73.2 	&	5143.6   &      73.5 &$|241)$	&	8684	& 77 	&	8687.8   &      76.3 \\
$|050)$	&	5156.2	& 0.054 &	5155.939 &      0.0544 &$|143)$	&	8802.3	& 90 	&	8801.8   &      88.6 \\
$|123)$	&	5275.7	& 32.1  &	5275.8   &      32.3 &	&	8921.4	& 90.9 	& 	8921.4   &      90.9 \\
$|221)$	&	5400.2	& 46.8 	&	5400.5   &      46.6 &$|062)$	&	8973.7	& 7.7 	&	8973.27  &      8.28 \\
	&	   	&	&	5460.9   &      114.4 &\\
\hline

\end{tabular}


\end{table}

\end{article}

\end{document}